\documentclass[aps,prd,twocolumn,superscriptaddress,showpacs]{revtex4}
\usepackage{graphicx}
\usepackage{dcolumn}
\usepackage{bm}
\usepackage{natbib}

\newcommand{\be}{\begin{equation}}
\newcommand{\ee}{\end{equation}}
\newcommand{\bea}{\begin{eqnarray}}
\newcommand{\eea}{\end{eqnarray}}

\newcommand{\nhat}{\hat{\bf n}}
\newcommand{\kvec}{{\bf k}}
\newcommand{\khat}{\hat{\bf k}}

\newcommand{\WMAP}{{\slshape WMAP~}}
\newcommand{\PLANCK}{{\slshape PLANCK~}}
\newcommand{\WMAPc}{{\slshape WMAP}}

\newcommand{\kpmask}{Kp0$\cap$S10$\setminus$ps~}

\begin{document}

\title{Correlating the CMB with Luminous Red Galaxies : The Integrated 
Sachs-Wolfe Effect}

\author{Nikhil Padmanabhan}
\email{npadmana@princeton.edu}
\affiliation{Joseph Henry Laboratories, Jadwin Hall, Princeton University, 
  Princeton, NJ 08544, USA}

\author{Christopher M. Hirata}
\affiliation{Joseph Henry Laboratories, Jadwin Hall, Princeton University,
  Princeton, NJ 08544, USA}

\author{Uro\v{s} Seljak}
\affiliation{Joseph Henry Laboratories, Jadwin Hall, Princeton University,
  Princeton, NJ 08544, USA}
\affiliation{ICTP, Strada Costiera 11, 34014 Trieste, Italy}

\author{David J. Schlegel}
\affiliation{Dept. of Astrophysical Sciences, Peyton Hall, Princeton University,
  Princeton, NJ 08544, USA}

\author{Jonathan Brinkmann}
\affiliation{Apache Point Observatory, 2001 Apache Point Road, Sunspot, New Mexico, 
88349-0059, USA}

\author{Donald P. Schneider}
\affiliation{Department of Astronomy and Astrophysics, Pennsylvania State University,
University Park, PA 16802}

\date{\today}

\begin{abstract}
We present a $2.5\sigma$ detection of the Integrated Sachs-Wolfe (ISW) effect
and discuss the constraints it places on cosmological parameters.
We cross-correlate microwave temperature maps from the \WMAP satellite 
with a 4000 deg$^{2}$ luminous red galaxy (LRG) overdensity map 
measured by the Sloan Digital Sky Survey.
These galaxies have accurate photometric redshifts ($\Delta z \sim 0.03$) and 
an approximately volume limited redshift distribution from 
$z\sim 0.2$ to $z\sim 0.6$
well suited to detecting the ISW effect. 
Accurate photometric redshifts allow us to perform a reliable
auto-correlation analysis of the LRGs, eliminating the uncertainty in the
galaxy bias, and combined with cross correlation signal, constrains cosmological parameters --
in particular, the matter density. 
We use a minimum variance power spectrum estimator that optimally weights the
data according to expected theoretical templates. 
We find a $2.5 \sigma$ signal in the Ka, Q, V, and W \WMAP bands,
after combining the information from multipoles $2 \le l < 400$. 
This  is consistent with the expected amplitude of the ISW effect, but 
requires a lower matter density than is usually assumed: the
amplitude, parametrized by the 
galaxy bias assuming $\Omega_M=0.3$, $\Omega_{\Lambda}=0.7$ and $\sigma_8=0.9$, is 
$b_{g} = 4.05 \pm 1.54$ for V band, with similar results for the other
bands. This should be compared to $b_g = 1.82 \pm 0.02$ from the auto-correlation analysis. 
These data provide only a weak confirmation ($2.5 \sigma$) of dark energy,
but provide a significant upper limit:
$\Omega_{\Lambda}=0.80_{-0.06}^{+0.03} (1\sigma) _{-0.19}^{+0.05} (2\sigma)$,
assuming a cosmology with
$\Omega_{M}+\Omega_{\Lambda}=1$, $\Omega_{b} = 0.05$, and $\sigma_{8}=0.9$, and
$w=-1$. The weak cross-correlation signal
rules out low matter density/high dark energy density
universes and, in combination with other data, 
strongly constrains models with $w<-1.3$. 
We provide a simple prescription to incorporate these constraints
into cosmological parameter estimation methods for $(\Omega_{M}, \sigma_{8},w)$.
We find no evidence for a systematic contamination of ISW signal, either
from Galactic or extragalactic sources, but we do detect some large 
statistical fluctuations on smaller scales that 
could affect analyses without the template weighting. 

\end{abstract}


\maketitle

\section{Introduction}
\label{sec:intro}

The cosmic microwave background (CMB) observed by the \WMAP satellite 
\cite{2003ApJ...583....1B} has 
been a font of information for cosmology. The positions
of the acoustic 
oscillations in the angular power spectrum of temperature 
fluctuations imprinted at the last scattering surface 
measures the curvature of the universe to unprecedented
accuracy. In addition, a careful modelling of the power
spectrum combined with other cosmological probes have allowed 
extremely precise measurements of the parameters of the
$\Lambda$CDM model. Subsequent measurements of the power spectrum
by \WMAP, as well as future experiments such as the \PLANCK satellite
will further constrain the space of 
cosmological models, as well as 
test our understanding
of the physics underlying recombination.

CMB temperature fluctuations are not just sourced by density 
fluctuations at the last scattering surface; anisotropies also
arise due to the interaction of photons with hot electrons in 
galaxies and clusters 
(the Sunyaev-Zeldovich \cite[SZ,][]{1980ARA&A..18..537S} 
and kinetic-SZ \cite{1980MNRAS.190..413S}
effects), as well as with the gravitational potentials 
along their propagation path (the Integrated 
Sachs-Wolfe \cite[ISW,][]{1967ApJ...147...73S} effect and
gravitational lensing). 

The ISW effect results from the red- (or blue-)shifting of CMB photons
as they propagate through gravitational potential wells. If these potentials
did not evolve, then the blueshift gained falling into a potential
well would exactly cancel the redshift emerging from the well;
evolving potentials spoil this cancellation. 
This is significant only on the largest scales, both because the 
power in the fluctuations of the potential is largest on large  scales,
and because integrating along the line of sight cancels out the 
temperature fluctuations (a photon is as likely to be redshifted 
as blueshifted). 

Although the above
effects are too small to unambigously detect in
current CMB data (for eg. \cite{2004astro.ph..4545H}), 
one can attempt to isolate them by cross-correlating
the temperature maps with suitable tracer populations 
\cite{2003ApJ...597L..89F,2004MNRAS.350L..37F, 2003astro.ph..7335S,
2004PhRvD..69h3524A,
2004Natur.427...45B, 2004ApJ...608...10N,2004astro.ph..4348B, 
2004astro.ph..6004H}.
This paper is a continuation of these efforts, attempting to
detect the ISW effect, but also goes one step further in making the 
ISW signal useful for cosmological parameter analysis. We combine the 
cross-correlation analysis with a galaxy auto-correlation measurement to 
remove the uncertainty due to galaxy bias. The auto-correlation 
analysis is only reliable when the 
galaxy population has well determined redshifts: we argue 
below that this is indeed the case for the sample used here. 
This paper does not focus on the auto-correlation analysis, 
so we perform a restricted analysis where we assume the shape of 
power spectrum and only constrain the amplitude. 
This is sufficient for the current purposes, as the errors are  
dominated by the statistical uncertainties in the ISW effect. A
detailed auto-correlation analysis will be presented in a
future paper. 

A galaxy catalog used for detecting the ISW effect
must satisfy several criteria. Since the ISW signal is 
only detectable on the largest scales, the catalog must cover
as large an area as possible. In addition, a large 
number of galaxies is necessary
to remove noise from
random Poisson fluctuations. Finally, to make theoretical 
interpretation possible, these galaxies must be drawn from a uniform
population with a well characterized redshift distribution.

Photometrically selected luminous red galaxies (LRGs) \cite{2004astro.ph..7594P}
from the Sloan Digital Sky Survey (SDSS)  \cite{2000AJ....120.1579Y} 
are a promising candidate. These galaxies
are amongst the most luminous in the Universe and so probe cosmologically
interesting volumes, including the transition from matter to dark
energy domination that sources the ISW signal. These galaxies are
old stellar populations with very uniform spectral energy distributions
that make uniform selection and accurate photometric redshifts possible.
Finally, the deep and wide-angle imaging of the sky by the SDSS allows
these galaxies to be selected over a large area with high densities.

Our goal is to extract the maximal amount of 
information on the ISW effect from the data. To do so, we employ 
quadratic estimator methods, which combine the data in an optimal 
way in the presence of noise and incomplete sky coverage
\cite{1997PhRvD..55.5895T,1998ApJ...506...64S}. In addition, we 
weight the information on different scales according to the 
expected scale dependence of the signal: 
the ISW effect is expected to dominate 
on large scales and be absent on small scales, so we weight these
different scales accordingly. We apply these techniques to the latest 
SDSS data reductions, ensuring the largest available 
SDSS sky coverage to date. 

Detecting an ISW effect would provide evidence of either 
a cosmological constant, quintessence or curvature independent of supernovae
observations of the luminosity distance \cite{2004ApJ...607..665R}. 
We investigate the cosmological implications of 
an ISW detection within the assumption of a flat universe, and in particular,
on its ability to constrain the matter density. The rapid increase of the ISW
signal with decreasing matter density allows us to place strong lower bounds 
on the matter density and suggests an alternative probe of the properties 
of the dark energy.

The paper is organized as follows : Sec.~\ref{sec:theory} introduces the 
ISW effect, developing the formalism to compute the predicted ISW signal. 
Sec.~\ref{sec:data} describes the \WMAP and SDSS data used in this paper.
Sec.~\ref{sec:estimate} and Sec.~\ref{sec:results} then discuss the 
cross correlation of these data, and the results obtained. Systematic effects
are discussed in Sec.~\ref{sec:systematics}, while cosmological implications
are considered in Sec.~\ref{sec:cosmology}. We conclude in Sec.~\ref{sec:discussion}.
Unless otherwise specified, we assume a $\Lambda$CDM cosmology with 
$\Omega_{M}=0.3$, $\Omega_{\Lambda} =0.7$, $H_{0} = 100 h~\rm{km/s/Mpc}$ and
$\sigma_{8}=0.9$. Where necessary, we use $h=0.7$.

\section{Theory}
\label{sec:theory}

We briefly review the ISW effect, and its cross-correlation
with the galaxy density 
(see also eg. \cite{1996PhRvL..76..575C,2002PhRvD..65j3510C,
2000ApJ...540..605P,2004PhRvD..69h3524A})
We consider linear functions of the density field projected
onto the sky, specializing to the cases of the ISW temperature perturbations 
and the galaxy overdensity field. Decomposing these projected fields into 
spherical harmonics allows us to compute the expected auto- and cross-power 
spectra. For simplicity, we restrict our discussion to flat universes 
($\Omega_{m} + \Omega_{\Lambda}=1$).

\subsection{Projections onto the sky}
\label{sec:theory_project}

We start with the 3D matter overdensity field, 
$\delta_{3D}({\bf y})$, assumed to be
an isotropic random variable. This allows us to define the  
3D matter power spectrum, $P(k)$, 
\be
\langle \delta_{3D}(\kvec) \delta_{3D}(\kvec') \rangle 
\equiv (2\pi)^{3} \delta(\kvec - \kvec') P(k) \,\,\,.
\label{eq:3dpowspec}
\ee
We project this density field onto the sky, 
\be
\rho(\nhat) = \int \, dy \, f(y) {\cal L} [\delta_{3D}(y,y\nhat)] \,\,\, ,
\ee
where $f(y)$ weights the density field as a function of comoving distance, $y$,
and ${\cal L}$ is a linear operator (independent of ${\bf y}$)
operating on the 3D density field. Simultaneously
expanding the projected field in spherical harmonics, $Y_{lm}(\nhat)$, and Fourier
transforming the 3D density field, we obtain
\be
\rho_{lm} = i^{l} \int \,\frac{d^{3} k}{2\pi^{2}} \int \, dy \, f(y)
   j_{l}(ky) Y_{lm}^{*}(\khat) {\cal L} [\delta_{3D}(y,\kvec)] \,\,\,,
\label{eq:rholm}
\ee
where we use the orthogonality of the $Y_{lm}$'s and the expansion 
of a Fourier wave in spherical coordinates,
\be
e^{-i\kvec \cdot \nhat y} = 4\pi \sum_{lm} \, i^{l} j_{l}(ky) 
Y_{lm}^{*}(\khat) Y_{lm}^{*}(\nhat) \,\,\, .
\ee

\subsubsection{Galaxy and Temperature Fluctuations}

Given the above equation, we specialize to the galaxy overdensity 
and the ISW temperature fluctuations. We assume 
that the galaxy overdensity, $\delta_{g,3D}$,
is related to the matter overdensity by a linear bias, $b_{g}$, 
\be
\delta_{g,3D} = {\cal L}_{g} [ \delta_{3D} ] = b_{g} \delta_{3D} \,\, ,
\ee
which implies that (Eq.~\ref{eq:rholm}),
\be
\delta_{g,lm} = i^{l} b_{g} \int \,\frac{d^{3} k}{2\pi^{2}} \int \, dy \, f(y)
   j_{l}(ky) Y_{lm}^{*}(\khat) \delta_{3D}(y,\kvec) \,\,\,.
\label{eq:galaxylm}
\ee
Although linear bias must break down at some scale, it appears to be 
well motivated both theoretically\cite{1998ApJ...504..607S}
and observationally \cite{2004ApJ...606..702T} on the 
large scales that the ISW effect is sensitive to. 
The redshift weighting $f(y)$ is simply given by the 
galaxy selection function, $\phi(y)$,
appropriately normalized,
\be
f(y) = \frac{y^2 \phi(y)}{\int \, dy\, y^{2} \phi(y)} \,\,.
\ee  
Working on large scales allows us to use 
linear perturbation theory where the growth of density fluctuations with 
time is separable,
\be
\delta_{3D}(y,\kvec) = D(y) \delta_{3D}(\kvec) \,\,,
\ee
where $D(y)$ is the growth factor. Substituting this above allows us
to write Eq. \ref{eq:galaxylm} in the following useful form,
\be
\delta_{g,lm} = i^{l} b_{g} \int \,\frac{d^{3} k}{2\pi^{2}} \, \delta_{3D}(\kvec)
Y_{lm}^{*}(\khat) W_{g}(k) \,\,,
\label{eq:galaxylm2}
\ee
where the window function $W_{g}(k)$ is given by
\be
W_{g}(k) = \int \, dy \, f(y) D(y) j_{l}(ky) \,\,.
\label{eq:galaxyw}
\ee

The temperature fluctuations due to the ISW effect are given by the
line of sight integral of the change in the gravitational potential 
to the last scattering surface,
\be
\left(\frac{\Delta T}{T}\right)_{ISW} = -2 \int_{0}^{y_{0}} \,dy \, 
\dot{\Phi}(y, y\nhat) \,\,\,,
\label{eq:ISWtempfluc}
\ee
where $\Phi$ is the gravitational potential, and the overdot is the 
derivative with respect to conformal distance (or equivalently, conformal
lookback time) at constant $\mathbf{y}_{3D}$. 
The gravitational potential can be related to
the 3D density fluctuations by Poisson's equation,
\be
\nabla^{2} \Phi = \frac{3}{2} H_{0}^{2} \Omega_{M} 
\frac{\delta_{3D}}{a} \,\,,
\label{eq:fisheq}
\ee
where $a$ is the scale factor, implying that,
\be
{\cal L}_{ISW} [ \delta_{3D} ] = - 3 H_{0}^{2} \Omega_{M} 
\frac{\partial}{\partial y} \left(\nabla^{-2} \frac{\delta_{3D}}{a} \right) \,\,.
\ee 
Taking the Fourier transform and substituting into Eq.~\ref{eq:rholm}, we find that,
\bea
\left(\frac{\Delta T}{T}\right)_{ISW,lm} = 
-i^{l} \int \,\frac{d^{3} k}{2\pi^{2}} 
\frac{3 \Omega_{M} H_{0}^{2}}{k^2} \int \, dy \, \nonumber \\
\times j_{l}(ky) Y_{lm}^{*}(\khat) \frac{\partial}{\partial y}
\left(\frac{\delta_{3D}(y,\kvec)}{a(y)}\right) \,\,\,.
\label{eq:ISWlm}
\eea
Again restricting to linear theory
allows us to write the above equation in the same form as Eq. 
\ref{eq:galaxylm2},
\be
\left(\frac{\Delta T}{T}\right)_{ISW,lm} = -i^{l} \int \,\frac{d^{3} k}{2\pi^{2}}
\, \delta_{3D}(\kvec) Y_{lm}^{*}(\khat) W_{ISW}(k) \,\,,
\ee
where the ISW window function is given by 
\be
W_{ISW}(k) = \frac{3 \Omega_{m} H_{0}^{2}}{k^{2}} \int_{0}^{y_{0}} \,dy\, 
j_{l}(ky) \partial_{y}{\left(\frac{D}{a}\right)}  \,\,.
\label{eq:isww}
\ee

It is straightforward to compute 
the relevant power spectra. The galaxy power spectrum, 
$C_{l}^{gg} \equiv \langle \delta_{g,lm} \delta^{*}_{g,lm} \rangle$, is given by,
\be
C_{l}^{gg} = 4\pi \int \,dk\, \frac{\Delta^{2}(k)}{k} | W_{g}(k) |^{2} \,\,\,,
\label{eq:galaxycl}
\ee
where we have used the definition of the three dimensional power spectrum 
(Eq. \ref{eq:3dpowspec}) and $\Delta^{2}(k)$ is the variance per logarithmic
wavenumber,
\be
\Delta^{2}(k) \equiv \frac{1}{(2\pi)^{3}} 4\pi k^{3} P(k) \,\,\,.
\ee
The cross correlation between the ISW temperature fluctuations and 
galaxy overdensities is similarly given by,
\be
C_{l}^{g-ISW} = 4\pi \int \,dk\, \frac{\Delta^{2}(k)}{k} W_{g}(k) W_{ISW}(k) \,\,\,.
\ee

\subsection{Predictions}

\begin{figure}
\includegraphics[width=3in]{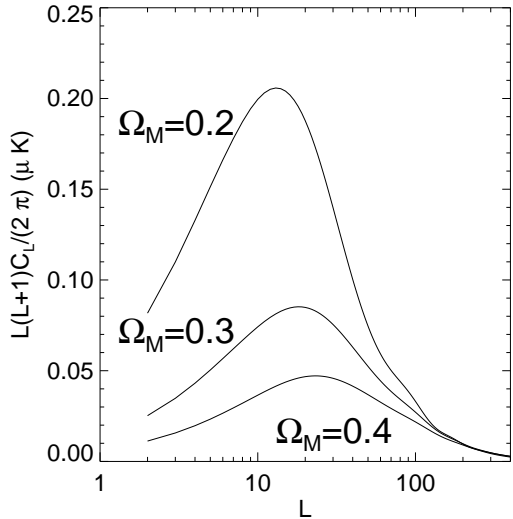}
\caption{\label{fig:isw_plottemplate} Predictions for the ISW signal given 
the redshift distribution of LRGs described in 
Sec.~\ref{sec:lrg}, $b_{g}=1$, $\sigma_{8}=0.9$, 
and a flat universe. The different curves show the effect of changing the matter
density. In particular, observe that the effect becomes stronger as the 
matter density decreases.}
\end{figure}

The above formalism allows us to compute the expected ISW power spectrum
given the redshift distribution of a tracer population and a particular
cosmology. Fig.~\ref{fig:isw_plottemplate} shows the ISW power spectrum
assuming the LRG redshift distribution introduced below (Sec.~\ref{sec:lrg})
and a flat cosmology. The growth factor is given by 
\be
D(a) = \frac{5}{2} \Omega_{M,0}\frac{H(a)}{H_{0}} \int^{a}_{0} 
\frac{da'}{\left(a' H(a')/H_{0}\right)^{3}} \,\,,
\ee
where $a \equiv 1/(1+z)$ is the scale factor, and $H(a)$ is determined
by the Friedmann equation.
We note that (within the context
of flat cosmologies), $D/a$ is independent of redshift if $\Omega_{M}=1$,
leading to the lore that the detection of the ISW effect is 
evidence for some form of dark energy, although we note that universes
with curvature and no dark energy/cosmological constant also lead to an 
ISW effect.

Finally, a computational note on performing the spherical bessel integrals 
of the previous section : we follow \cite{2000MNRAS.312..257H} and recast these
as logarithmically discretized Hankel transforms. 
In this form, the integrals can 
be efficiently performed via FFT convolutions using the FFTLOG algorithm
\cite{talman_fftlog} as implemented in \cite{2000MNRAS.312..257H}.

\section{Data}
\label{sec:data}

\subsection{CMB temperature from \WMAPc}
\label{sec:wmap}

The \WMAP mission \cite{2003ApJ...583....1B} is designed to produce
all-sky maps of the CMB at multipoles up to $l\sim$several hundred.  This
analysis uses the first public release of \WMAP data, consisting of the first
year of observations from the Sun-Earth L2 Lagrange point.  \WMAP carries
ten differencing assemblies (DAs), each of which measures the difference
in intensity of the CMB at two points on the sky; a CMB map is constructed
from these temperature differences as the satellite rotates. 
\WMAP observes in the K (1 DA), Ka (1 DA), Q (2 DAs), V (2 DAs),
and W (4 DAs) bands corresponding to central frequencies of 
23, 33, 41, 61, and 94 GHz, respectively.
The \WMAP team has pixelized the data from each DA in the
HEALPix \footnote{URL: {\tt http://www.eso.org/science/healpix/}}
pixelization system at resolution 9 \cite{2003ApJS..148....1B,
2003ApJS..148...63H}.  This system has 3,145,728 pixels, each 
47.2 sq. arcmin in area. These maps are not beam-deconvolved; this, combined
with the \WMAP scan strategy, results in nearly uncorrelated Gaussian
uncertainties on the temperature in each pixel.

We limit this analysis to the Ka through W bands. The K band is heavily
contaminated by Galactic emission, increasing the number of possible
systematic effects and making any error analysis unreliable. We apply the
Kp0 mask specified by the \WMAP team to mask out regions where Galactic
foregrounds dominate. In addition, we reject pixels in the \WMAP point source
mask, including a 0.6 degree exclusion radius around each source. Finally,
since the SDSS area used only covers about a tenth of the sky, we eliminate 
all pixels greater than 10 degrees from the SDSS mask, allowing us to speed
up our cross-correlation analysis. This mask, denoted \kpmask, 
admits 756,078 HEALPix pixels
(9915 sq. deg). The variance in each of these pixels is computed directly from
the number of observations of each pixel, as specified by the \WMAP team.

We chose not to use either the \WMAP ``Internal Linear Combination''
(ILC) map \citep{2003ApJS..148....1B}, or the foreground cleaned map of 
\citep{2003PhRvD..68l3523T}, to avoid a number of practical difficulties.
These maps lose the frequency dependence of the original maps (useful to
identify contaminating signals) and have complicated pixel-pixel noise
correlations (increasing the complexity of the error analysis).

\subsection{SDSS Luminous Red Galaxies}
\label{sec:lrg}

\begin{figure}
\includegraphics[width=3in]{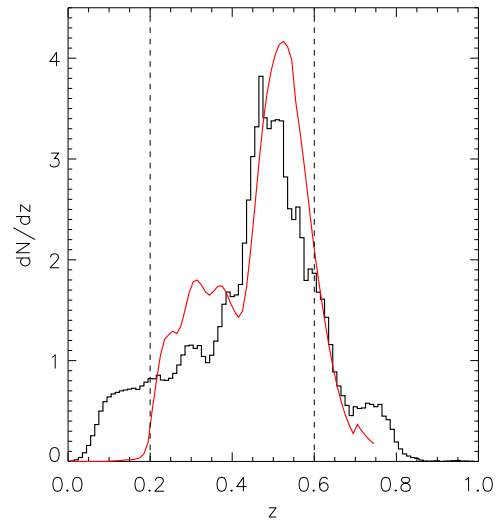}
\caption{\label{fig:lrgdndz}The LRG redshift distribution.  The 
histogram shows the photometric redshift distribution, the curve is the true 
redshift distribution estimated by regularized deconvolution of the 
photo-$z$ errors. The dotted lines show the photometric redshift cuts imposed 
at $z=0.2$ and $0.6$.}
\end{figure}

The Sloan Digital Sky Survey \cite{2000AJ....120.1579Y} is an
ongoing effort to image approximately $\pi$ steradians of the sky, and
obtain spectra of approximately one million of the detected objects
\cite{2002AJ....124.1810S, 2001AJ....122.2267E}. The
imaging is carried out by drift-scanning the sky in photometric conditions
\cite{2001AJ....122.2129H}, in five bands ($ugriz$)
\cite{1996AJ....111.1748F, 2002AJ....123.2121S} using a specially designed
wide-field camera \cite{1998AJ....116.3040G}. 
Using these data, objects are targeted for spectroscopy 
\cite{2003AJ....125.2276B}
and are observed with a 640-fiber spectrograph on the same telescope. All
of these data are processed by completely automated pipelines that detect
and measure photometric properties of objects, and astrometrically
calibrate the data \cite{2001adass..10..269L, 2003AJ....125.1559P}. The
SDSS is nearing completion, and has had four major data releases
\cite{2002AJ....123..485S, 2003AJ....126.2081A, 2004AJ....128..502A}
\endnote{URL: \texttt{www.sdss.org/dr3}}. 
This paper uses all data observed through Fall 2003, processed and
calibrated as described in \cite{oriondatarelease}.

The usefulness of LRGs as a cosmological
probe has been appreciated by a number of authors
\cite{2000AJ....120.2148G, 2001AJ....122.2267E}. These are typically the
most luminous galaxies in the universe, and therefore probe cosmologically
interesting volumes.  In addition, these galaxies are generically old
stellar systems with uniform spectral energy distributions
(SEDs) characterized principally by a strong 
discontinuity at 4000~\AA. This
combination of a uniform SED and a 
strong 4000~\AA~break make LRGs an ideal candidate for
photometric redshift algorithms, with redshift accuracies of $\sigma_z
\sim 0.03$ \cite{2004astro.ph..7594P}. We briefly outline the construction of the
photometric LRG sample used in this paper below; a detailed
discussion of the selection criteria and properties of the sample
is in \cite{2004astro.ph..7594P, padauto}.

Our selection criteria are based on the spectroscopic 
selection of LRGs described in \cite{2001AJ....122.2267E},
extended to lower apparent luminosities. We
select LRGs by choosing galaxies that both have colors consistent with
an old stellar population, as well as absolute luminosities greater
than a chosen threshold. The first criterion is simple to implement since
the uniform SEDs of LRGs imply that they lie on an extremely tight locus
in the space of galaxy colors; we simply select all galaxies that lie
close to that locus. More specifically, we can define three (not 
independent) colors that describe this locus,
\bea
c_{\perp} &\equiv & (r-i) - 0.25(g-r) - 0.18 \,\,\, , \nonumber \\
d_{\perp} &\equiv & (r-i) - 0.125(g-r) \,\,\,, \nonumber \\
c_{||} &\equiv & 0.7 (g-r) + 1.2(r-i-0.18) \,\,\,,
\label{eq:perpdef}
\eea
where $g$, $r$, and $i$ are the SDSS model magnitudes 
\cite{2002AJ....123..485S} in these bands respectively.
We now make the following color selections,
\bea
{\rm Cut\,\,I :} & \mid c_{\perp} \mid < 0.2; \nonumber \\
{\rm Cut\,\,II :} & d_{\perp} > 0.55, \,\,\, g-r > 1.4.
\label{eq:colourcuts}
\eea
Making two cuts (Cut I and Cut II) is convenient since the LRG
color locus changes direction sharply as the 4000 \AA~break
redshifts from the $g$ to the $r$ band; this division divides the
sample into low redshift (Cut I, $z < 0.4$) and high
redshift (Cut II, $z > 0.4$) samples.

In order to implement the absolute magnitude cut, we follow
\cite{2001AJ....122.2267E} and impose a cut in the
galaxy color-magnitude space. The specific cuts we use are
\bea
{\rm Cut\,\,I :} && r_{Petro} < 13.6 + \frac{c_{||}}{0.3}, \,\,\,
r_{Petro} < 19.7,
\nonumber \\
{\rm Cut\,\,II :} && i < 18.3 + 2d_{\perp}, \,\,\, i < 20,
\eea
where $r_{Petro}$ is the SDSS $r$ band Petrosian magnitude 
\cite{2002AJ....123..485S}.  Finally, we reject all objects that resemble 
the point-spread function, or if they have colors 
inconsistent with normal galaxies; these cuts attempt to remove 
interloping stars.

\begin{figure}
\includegraphics[width=3in]{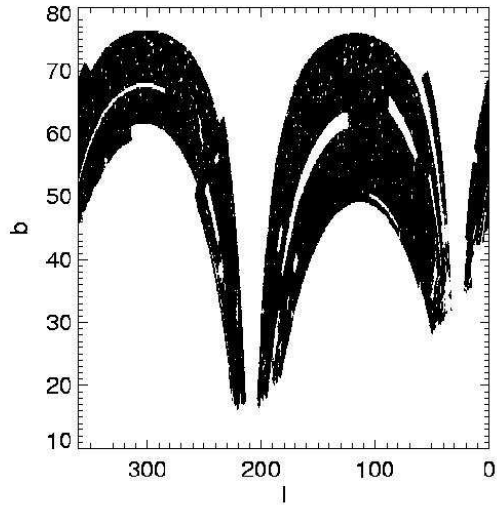}
\caption{\label{fig:lrgmask}The LRG angular distribution in 
Galactic coordinates. The gaps in the distribution are due
to the stellar mask, nonphotometric data, and Galactic 
extinction.}
\end{figure}

Applying these selection criteria to the $\sim $ 5500 degress of
photometric SDSS imaging in the Galactic North yields a catalog of
approximately 900,000 galaxies. Applying the simple template fitting
photometric redshift algorithm of \cite{2004astro.ph..7594P}, we restrict this catalog
to galaxies with $0.2 < z_{photo} < 0.6$, leaving us with $\sim$ 650,000
galaxies. We use the regularized inversion method of \cite{2004astro.ph..7594P} as
well as the photometric redshift error distribution presented there, to
estimate the true redshift distribution of the sample. The results,
comparing the photometric and true redshift distributions are shown in
Fig.~\ref{fig:lrgdndz}. The LRGs are approximately volume limited from $z\sim 0.2$
to $z\sim 0.55$. Comparing this to the S/N estimates of \cite{2004astro.ph..1166A},
we see that the LRG redshift distribution is well suited to detecting the ISW effect.
Therefore, despite the availability of accurate photometric redshifts, 
we do not weight the LRGs in redshift 
any differently than they already are in the sample we use here.

This catalog is pixelized as a number
overdensity, $\delta_g=\delta n/\bar n$, onto a HEALPix pixelization of the
sphere, with 3,145,728 pixels. Since the photometric catalogs are
incomplete around bright stars, we mask regions around the stars in the
Tycho astrometric catalog \cite{2000A&A...355L..27H}. 
We also exclude data from the three southern SDSS stripes due to
difficulties in photometrically calibrating them relative to the data in the 
Northern Galactic Cap.
The final catalog covers an
solid angle of 3,893 square degrees (296,872 HEALPix resolution 9 pixels)  
and contains 503,944 galaxies at a mean density of $1.70$ galaxies per
pixel. The sky coverage is shown in Fig.~\ref{fig:lrgmask}.

\section{Estimating the Cross Correlation}
\label{sec:estimate}

We start by organizing the temperature fluctuations
and the galaxy overdensities into a single data vector,
\be
\mathbf{x} = (\mathbf{x}_{B,T},\mathbf{x}_{g}) \,\,,
\ee
where $\mathbf{x}_{B,T}$ is a vector with the measured 
CMB temperature (with the monopole and dipole subtracted) 
in band $B$ at every HEALPix pixel;
analogously, $\mathbf{x}_{g}$ is the LRG number overdensity.
We suppress the band
subscript for simplicity, with the implicit understanding
that we always refer to the cross correlation of 
a single \WMAP band with the LRG overdensity. 
The covariance matrix of $\mathbf{x}$ is,
\be
\mathbf{C} = \mathbf{C}_{diag} + \left(
\begin{array}{cc} & C^{gT} \\ C^{gT} & \end{array}\right) \,\,,
\label{eq:cdef}
\ee
where $\mathbf{C}_{diag}$ is given by,
\be
\mathbf{C}_{diag} = \left(\begin{array}{cc} C^{TT}+N^{TT} & \\ & C^{gg} + N^{gg} 
\end{array}\right) \,\,,
\label{eq:cdiagdef}
\ee
where $N^{xx}$ is the pixel noise matrix.
The submatrices $C^{TT}, C^{gg}$ and $C^{gT}$ are defined by
\be
C^{ab}_{ij} = \sum_{lm} C^{ab}_{l} 
Y_{lm}^{*}(\hat{n}_{i}^{a}) Y_{lm}(\hat{n}_{j}^{b}) \,\,,
\ee
where $\hat{n}_{i}^{a}$ is the position (on the sky) of the 
$i^{th}$ point of the vector $\mathbf{x}_{a}$. The temperature-temperature,
galaxy-galaxy and galaxy-temperature angular power spectra are denoted 
by $C^{TT}_{l}, C^{gg}_{l}$ and $C^{gT}_{l}$ respectively. In our analysis 
below, we use the best fit $\Lambda$CDM prediction for the temperature
angular power spectrum, multiplied by the appropriate WMAP instrumental response
at each multipole. The galaxy power spectrum is estimated using a pseudo-$C_{l}$
estimator \citep{padauto}, and fit by the non-linear power spectrum of
\cite{2003MNRAS.341.1311S}, multiplied by a constant linear bias. We project out
the monopole and dipole of both these power spectra by setting the power in 
the $l=0,1$ modes to a value ($10^{-2}$) much greater than the true 
power spectrum. In addition, the measured galaxy-galaxy power spectrum has
excess power on large scales $(l < 20)$ due to calibration errors; we boost
the prior power spectrum to account for this.

We parametrize $C^{gT}_{l}$ as a sum of bandpowers, 
$\tilde{P}_{i,l}$, with amplitudes $c_{i}$ to be estimated,
\be
C^{gT}_{l} = \sum_{i} c_{i} \tilde{P}_{i,l} \,\,.
\ee
We consider two bandpowers in this paper. The first are
``flat'' bandpowers given by
\bea
\tilde{P}_{i,l}  & = & B(l) \,;\, l_{i,min} \leq l <  l_{i,max} \nonumber \\
& = & 0 \,;\, \rm{otherwise} \,\,,
\eea 
where $B(l)$ is the WMAP instrumental 
response\cite{2003ApJS..148...39P}. This parametrizes the power spectrum 
as a sum of step functions and is useful
when the shape of the power spectrum is unknown. However, as 
discussed in Sec.~\ref{sec:theory}, the shape of the 
ISW component of the galaxy-temperature correlation is well determined
by the cosmology and the redshift distribution of the galaxies with
only its amplitude (equivalently, the galaxy bias) unknown. We define the 
``template'' bandpowers by replacing the flat bandpowers for $l < l_{max}$
with the shape of the ISW correlation, i.e.
\bea
\tilde{P}_{1,l} & = & B(l) C^{g-ISW}_{l}\,;\,  2 \leq l <  l_{max}  \nonumber \\
& = & 0 \,;\, \rm{otherwise} \,\,.
\eea
This definition implies that $c_{1}$ directly measures the galaxy bias by 
optimally combining information from multipoles $< l_{max}$.

We are not restricted to estimating $C^{gT}_{l}$,
but can simultaneously measure $C^{TT}_{l}$ and $C^{gg}_{l}$. Since doing 
so is not the goal of this paper, we simply measure the amplitude 
of the input $C^{TT}_{l}$ and $C^{gg}_{l}$ analogous to the template 
bandpower above. This provides a useful runtime check of our pipeline,
as well as allowing us to estimate the galaxy bias.

We estimate the $c_{i}$ by forming quadratic combinations of the data 
\cite{1997PhRvD..55.5895T, 1998ApJ...506...64S},
\be
q_{i} = \frac{1}{2} \mathbf{x}^{t} \mathbf{C}_{diag}^{-1} 
\frac{\partial \mathbf{C}}{\partial c_{i}} \mathbf{C}_{diag}^{-1} \mathbf{x} \,\,.
\ee
These are related to the estimated $\hat{c}_{i}$ by the Fisher matrix, $\mathbf{F}$,
\be
\hat{c}_{i} = \sum_{j} (\mathbf{F}^{-1})_{ij} q_{j} \,\,,
\ee
where 
\be
\mathbf{F}_{ij} = \frac{1}{2} \rm{tr}\left[ 
\mathbf{C}_{diag}^{-1} \frac{\partial \mathbf{C}}{\partial c_{i}}
\mathbf{C}_{diag}^{-1} \frac{\partial \mathbf{C}}{\partial c_{j}} \right] \,\,.
\ee
If $C^{gT}_{l} \ll \sqrt{C^{gg}_{l} C^{TT}_{l}}$, then the $\hat{c}_{i}$ are a
good approximation to the maximum likelihood estimates of the $c_{i}$. 
The covariance matrix of the $\hat{c}_{i}$ is 
the inverse of the Fisher matrix, if the 
fiducial power spectra and noise used to compute $\mathbf{C}_{diag}^{-1}$ correctly
describe the data. These assumptions must 
be tested and calibrated with simulations (Sec.~\ref{sec:sims}).

Implementing the above algorithm is complicated by the sizes of the datasets; the data vector
has 1,052,950 elements making both storing and naively manipulating the covariance 
matrix impossible on presently available computers. 
Working in harmonic space is also not possible 
due to the complicated geometry of the SDSS and WMAP masks. We implement
the methods of \citep{2003NewA....8..581P} extended to the sphere. All
matrix-vector operations are performed using convolutions using the 
spherical harmonic transform code of \cite{2004astro.ph..6004H}, 
which scales as $N^{3/2}$,
compared to the $N^{2}$ scaling of direct matrix multiplications. Matrix inversions are
performed with a preconditioned conjugate gradient code \citep{1992nrfa.book.....P}, 
using the preconditioner in Appendix B of \cite{2004astro.ph..6004H}; this typically
converges to a fractional precision (with an $L_{2}$ norm) of $10^{-8}$ 
in $\sim 100$ iterations. Finally, the Fisher matrix is computed using 
a $Z_{2}$ stochastic trace algorithm \citep{2003NewA....8..581P}; approximately
25 random vectors achieves the necessary precision.

\section{Results}
\label{sec:results}

\begin{figure}
\includegraphics[width=3in]{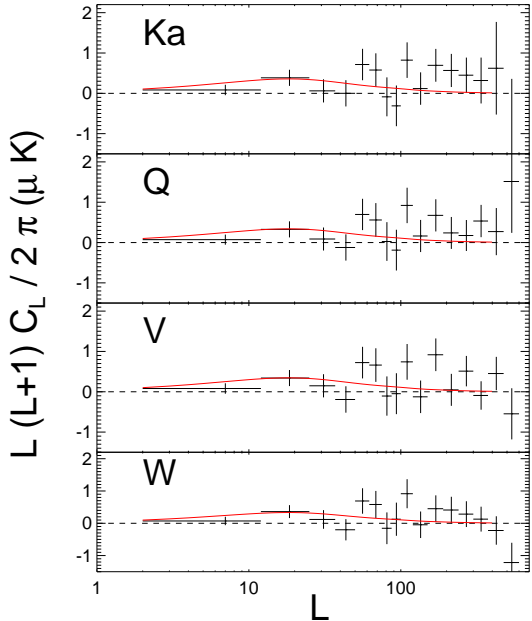}
\caption{\label{fig:isw_flatpower} The cross correlation between WMAP and the SDSS LRG 
density, as measured in flat bandpowers for the four WMAP bands considered in the paper.
Also plotted is predicted ISW signal for a $\Omega_{M}=0.3, \Omega_{\Lambda}=0.7$ 
universe, scaled by the measured galaxy bias in Table~\ref{tab:bias}. Note that these 
power spectra have been beam-deconvolved.}
\end{figure}

\begin{table}
\begin{tabular}{ccc}
\hline\hline
Band & $b_{g}$ ($l<50$) & $b_{g}$ ($l<400$) \\
\hline
Ka & $3.12 \pm 1.65$ & $4.20 \pm 1.56$ \\
Q  & $2.77 \pm 1.65$ & $3.99 \pm 1.56$ \\
V  & $2.89 \pm 1.65$ & $4.05 \pm 1.56$ \\
W  & $2.81 \pm 1.66$ & $3.91 \pm 1.57$ \\
V (flipped) & - & $0.39 \pm 1.58$ \\
V (inverted) & - & $0.74 \pm 1.57$ \\
\hline
\end{tabular}
\caption{\label{tab:bias} The measured galaxy bias as estimated
from the ISW signal, for templates truncated at $l_{max}=50$
and $l_{max}=400$. The errors
on the data are computed from the Fisher matrix.
Note that the different bands are correlated on the largest scales, explaining
the similarities in the values of bias measured. Also shown are the measured 
biases for a flipped and inverted V band map; these maps are not expected 
to be correlated with the LRG density.} 
\end{table}

The results of cross-correlating the WMAP temperature maps with the 
the LRG density are shown in Fig.~\ref{fig:isw_flatpower}, where the power 
spectrum has been estimated in flat bandpowers. For $l < 50$ where the 
ISW signal is expected to be the strongest (Fig.~\ref{fig:isw_plottemplate}),
we see evidence for a signal, peaking around $l \sim 20$. This signal
is seen in all four bands and appears to be independent of frequency, consistent
with the achromatic ISW effect.

We want to estimate the amplitude of this signal, parametrized by the LRG bias. 
Performing a $\chi^{2}$ fit to the above cross power 
spectra is undesirable for two reasons -- (i) the flat bandpowers
do not optimally use the information from the individual multipoles, 
and (ii) inverting a noisy covariance matrix (due to the stochastic
trace estimation) is likely to incorrectly estimate the 
significance of the detection. To avoid both these problems, we 
use the template bandpowers described in the previous section to estimate
the amplitude of the expected ISW signal (assuming our fiducial 
cosmology). We consider two values of $l_{max}$ at which we truncate the 
ISW templates, $l_{max}=50$, and $l_{max}=400$. Truncating the template
at large $l$ ensures that we optimally use all the information present in 
the multipoles. However, we do see (Fig.~\ref{fig:isw_flatpower}) some 
large fluctuations at small scales, $l > 50$. As we discuss in the next section,
these appear to be statistical fluctuations, but 
choosing $l_{max}=50$ allows us to be conservative and ignore these multipoles.

The results are in Table~\ref{tab:bias}, and the best fit
templates are plotted in Fig.~\ref{fig:isw_flatpower}. 
The signal again appears to be 
achromatic as expected. The values of the bias are
consistent with measurements of the bias from the galaxy-galaxy
power spectrum, $b_{g} = 1.82 \pm 0.02$, estimated as discussed in the
previous section. We also note the values of the bias obtained
from truncating the templates at $l_{max}=50$ and $l_{max}=400$ are consistent,
although, not surprisingly, the error associated with $l_{max}=50$ is 
higher.

Table~\ref{tab:bias} also presents the amplitude of the cross-correlation between 
the LRG density and flipped (north and south reversed) and inverted (antipodal
points exchanged) V band \WMAP maps. These maps should not be correlated with the 
LRG density, and provide a useful systematic test internal to the data. Both biases
obtained are consistent with zero as expected.

\section{Systematics}
\label{sec:systematics}

\subsection{Pipeline Simulations}
\label{sec:sims}

\begin{table*}
\begin{tabular}{cccccc}
\hline\hline
Band & $b_{g}$ ($l<50$) & $b_{g}$ ($l<400$) 
  & $\chi^{2}$ (flat) & $\chi^{2}$ ($l_{max}<50$) & $\chi^{2}$ ($l_{max}<400$) \\
& & & (16 dof) & (8 dof) & (2 dof) \\
\hline
Ka & $0.30 \pm 1.64$ & $0.30 \pm 1.56$ & 16.88 (0.39) & 9.90 (0.27) & 4.21 (0.12) \\
Q  & $0.30 \pm 1.64$ & $0.29 \pm 1.56$ & 19.29 (0.25) & 8.73 (0.37) & 5.78 (0.06) \\
V  & $0.29 \pm 1.65$ & $0.30 \pm 1.57$ & 15.00 (0.52) & 7.64 (0.47) & 3.99 (0.14) \\
W  & $0.28 \pm 1.69$ & $0.25 \pm 1.60$ & 20.82 (0.19) & 8.51 (0.38) & 4.10 (0.13) \\ 
\hline
\end{tabular}
\caption{\label{tab:sims} The average galaxy bias derived from the 110 \WMAP simulations
cross-correlated with the actual LRG overdensity map,
for $l_{max}=50$ and $l_{max}=400$. Recall that the simulated microwave maps are
uncorrelated with actual LRG map, but that the different \WMAP frequencies are
correlated. Also shown is the $\chi^{2}$ of the average power spectrum compared
with zero and (in parentheses) the probability that $\chi^{2}$ would
be greater than the measured value for a correct model. Note that we also
show the $\chi^{2}$ value for the flat bandpowers in Fig.~\ref{fig:isw_flatpower}.}
\end{table*}

End-to-end simulations are essential both to validate the pipeline, as well
as to calibrate errors obtained from the Fisher matrix. Since realistically
simulating the galaxy population would involve understanding the formation
of LRGs and therefore, is not currently feasible, we simulate the microwave
sky and cross-correlate these realizations with the actual LRG density map. 
Since the ISW cross-correlation is much smaller than the
individual auto-power spectra, we simply generate microwave sky maps
that are uncorrelated with the LRGs.

The primary microwave temperature fluctuations are simulated by generating
Gaussian random fields with an angular power spectrum identical to the 
best-fit $\Lambda$CDM spectrum determined by \WMAP \cite{2003ApJS..148..175S}. 
These maps
are then convolved with the appropriate 
instrument beams\cite{2003ApJS..148...39P}; we assume circular
beams for simplicity. Note that the temperature fluctuations are correlated
between the different frequency bands, allowing us to estimate the 
frequency correlation of the power spectrum estimates. We add
noise by using the 110 simulated noise maps 
provided by the \WMAP team \cite{2003ApJS..148...63H}.
In addition to white noise, these maps also simulate the $1/f$ detector noise and 
inter-detector correlations \cite{2003ApJS..148...29J}. We make no attempt to 
add either Galactic or extragalactic foregrounds to these maps; these
are small (especially for the V and W bands) outside of the Kp0 mask\cite{2003ApJS..148...97B}. 
All these maps
are then masked with the \kpmask mask, and then correlated with the 
actual LRG density in an identical manner to the actual temperature data.

The results from the 110 simulations are summarized in 
Table~\ref{tab:sims}. The frequency correlations on large scales are evident
from the identical values of the average bias obtained. Note that
measured power spectra are consistent with zero, both for the flat and
template bandpowers, with $\chi^{2}/\rm{dof} \sim 1$ in all cases.
Finally, we observe that the error on the bias from the Fisher matrix 
is consistent with the run-to-run simulation error (within the variance
of the simulations, $\Delta \sigma/\sigma \sim 1/2\sqrt{55}$).

\subsection{Galactic Foregrounds}
\label{sec:foreg}

The two sources of contamination in the galaxy catalog that could correlate
with Galactic microwave emission are stellar contamination and incorrect 
Galactic extinction corrections. The \kpmask mask by construction excludes
regions with the worst contamination; we test for 
any residual contamination
by directly estimating the level of contamination (Sec.~\ref{sec:foreg1}) and
cross-correlating foreground emission templates with the galaxy density 
(Sec.~\ref{sec:foreg2}).

\subsubsection{Stellar density and Galactic extinction contamination}
\label{sec:foreg1}

In order to estimate the level of stellar contamination and
incorrect extinction corrections, we compute the zero-lag 
correlation between the LRG density maps and maps of the stellar
density and Galactic extinction. To avoid discretization
effects, we re-pixelize the LRG catalog onto a HEALPIX sphere with
49152 (resolution 6) pixels. We estimate the stellar density by
selecting stars identified by the SDSS photometric pipeline, with $r$ band
PSF magnitudes between $18.0$ and $19.5$, and pixelizing these onto the
same HEALPIX sphere. The extinction map is constructed directly from
the $E(B-V)$ reddening map of \cite{1998ApJ...500..525S}.

\begin{figure}
\includegraphics[width=3in]{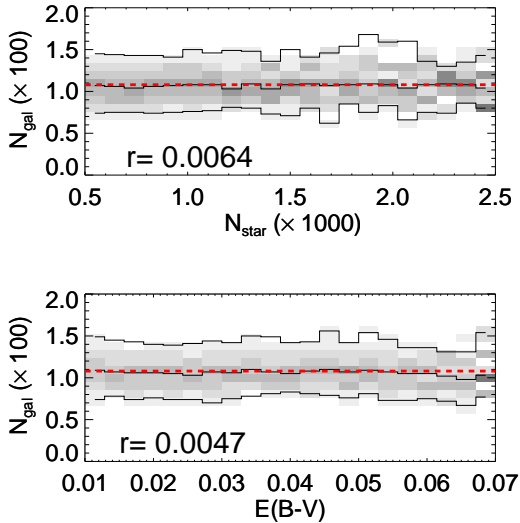}
\caption{\label{fig:LRG_star} Conditional contour plots of the
galaxy density as a function of the stellar density and galactic extinction,
$E(B-V)$. The contours are the 5\%, 50\% and 95\% contours, while the horizontal
line is the mean galaxy density.
Also shown are the correlation coefficients, $r$ for both datasets. There is no
evidence for significant stellar contamination or incorrect extinction corrections
in these data.}
\end{figure}

Contour plots of the LRG density as a function of stellar density and 
Galactic extinction are in Fig.~\ref{fig:LRG_star}; the contours show
conditional probabilities of 5\%, 50\% and 95\%. The near-horizontal 
contours (even at the extremes) suggest negligible levels of contamination.
The cross-correlation coefficients of $r=0.006$ (stellar density) 
and $r=0.004$ (extinction) further support this conclusion.

\subsubsection{Foreground Templates} 
\label{sec:foreg2}

\begin{figure}
\includegraphics[width=3in]{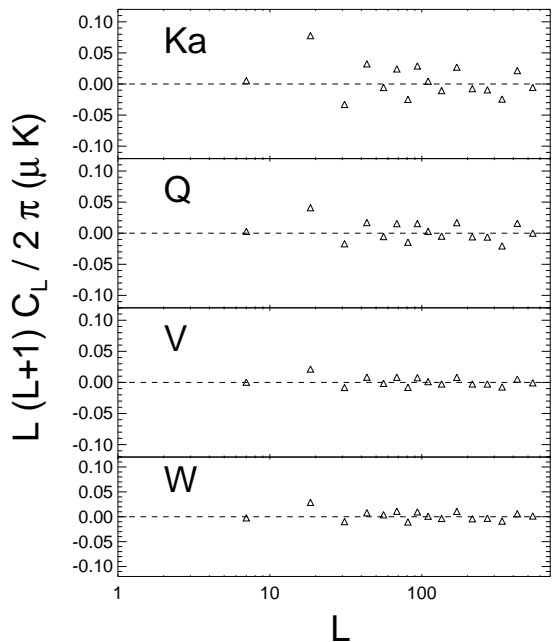}
\caption{\label{fig:isw_fgplot} The cross correlation between the \WMAP foreground
templates and the SDSS LRG density, measured in the same bandpowers as Fig.~\ref{fig:isw_flatpower}.
The contamination from known foregrounds is 
clearly subdominant to the CMB temperature cross correlation.}
\end{figure}

A second test of Galactic foreground contamination is to cross-correlate 
templates of Galactic emission 
with the LRG density maps. We construct these templates
with three components:
\begin{itemize}
\item Thermal dust emission : We use the 100 micron maps of dust emission
\cite{1998ApJ...500..525S}, extrapolated to the \WMAP frequencies using the
two-component dust model (Model 8) of \cite{1999ApJ...524..867F}.
\item Free-Free emission : We model the free-free component with the 
all sky $H\alpha$ maps of \cite{2003ApJS..146..407F} using the prescription 
in \cite{2003ApJS..148...97B}.
\item Spinning Dust/Synchrotron Emission:
\WMAP shows evidence for an additional source of emission at low frequencies 
due either to synchrotron emission \cite{2003ApJS..148...97B}, or spinning
dust \cite{2003astro.ph.11547F}. Since the source of this emission is still 
under debate, we adopt the simple phenomenological prescription in 
\cite{2003astro.ph.11547F} to model this component.
\end{itemize}
These maps are constructed in the same pixelization as the \WMAP temperature
maps, and are masked with the \kpmask mask. The resulting maps 
(identical in form to the temperature maps) are then analyzed
to compute the cross-correlation power spectrum.

The results are shown in Fig.~\ref{fig:isw_fgplot}. As expected, the
contamination increases with decreasing frequency, with Ka showing
the worst contamination and V and W with negligible contamination. However,
Galactic foreground contamination is no more than 20\% of the
detected signal for Ka, and is less 
than 10\% in the other bands. Therefore, while Galactic foregrounds may
be responsible for the slightly 
higher value of the bias in the Ka band (Table~\ref{tab:bias}), 
they cannot explain
most of the signal in the temperature-galaxy correlation.
Unlike the ISW effect, extragalactic foregrounds, particularly 
point sources, are expected to show a dependence on frequency.
Table~\ref{tab:bias} and Fig.~\ref{fig:isw_flatpower}
suggest that the any such frequency dependence
is much less than the errors on the measurements. 

\subsection{Redshift distribution uncertainties}

\begin{figure}
\includegraphics[width=3in]{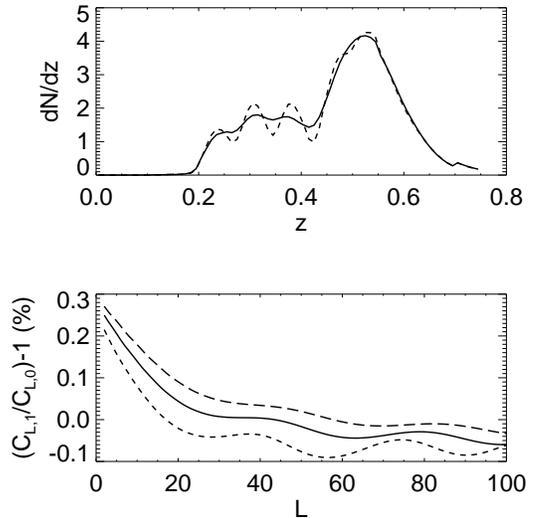}
\caption{\label{fig:isw_zdistdep} The upper panel 
shows the redshift distribution inferred from two different levels of 
regularization. The solid line is the distribution used throughout this paper, 
while the dashed line is less regularized by a factor of 10. 
The lower panel compares the templates
for $\Omega_{M}=0.2,0.3,0.4$ [short dashed, solid, long dashed] universes using
the above redshift distributions.}
\end{figure}

Predicting the ISW signal requires knowing the redshift distribution of the
LRGs. As discussed in \cite{2004astro.ph..7594P}, estimating the redshift distribution 
from measured photometric redshifts involves deconvolving the effect of 
photometric redshift errors. This process 
is unstable and must be regularized; unfortunately
the choice of the regularization parameter and therefore, the 
estimated redshift distribution, is not unique.
Fig.~\ref{fig:isw_zdistdep} estimates the error that this uncertainty introduces 
into the measurement of the bias. The upper panel compares the redshift distribution
used throughout this paper with a distribution estimated with a smaller regularization
parameter. The principal effect of reducing the regularization is ringing
in the inversion, caused by noise in the photometric redshift distribution. However,
as is evident from the lower panel, changing the redshift distribution makes only
sub-percent changes to the ISW signal. This insensitivity derives from the
fact that the greatest contribution to the ISW signal 
is coming from $z \sim 0.5$, where the inversion is most
stable. In addition, integrating along the line of sight makes
the signal principally sensitive to broadband features in the 
redshift distribution.

\subsection{Excess power at $l\sim 50,100, 200$}

\begin{figure}
\includegraphics[width=3in]{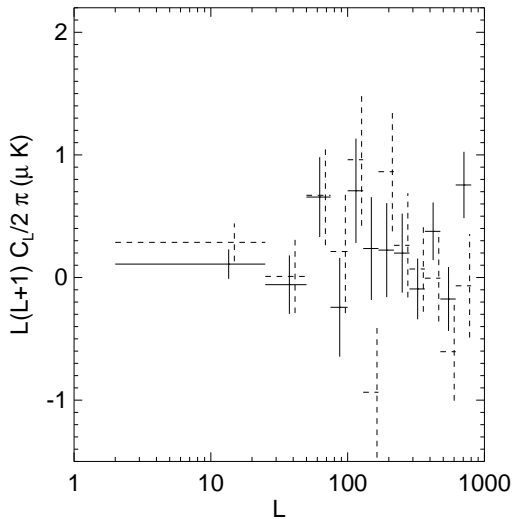}
\caption{\label{fig:iswcaps} The cross correlation between 
the \WMAP V band and the LRG density, measured
for the two contiguous 
SDSS regions separately. The measured power spectra are consistent within the
estimated errors. Note that the bands are 
$\sim$ twice the size of bands in Fig.~\ref{fig:isw_flatpower}, and the
dashed points have been displaced for clarity.}
\end{figure}

\begin{figure}
\includegraphics[width=3in]{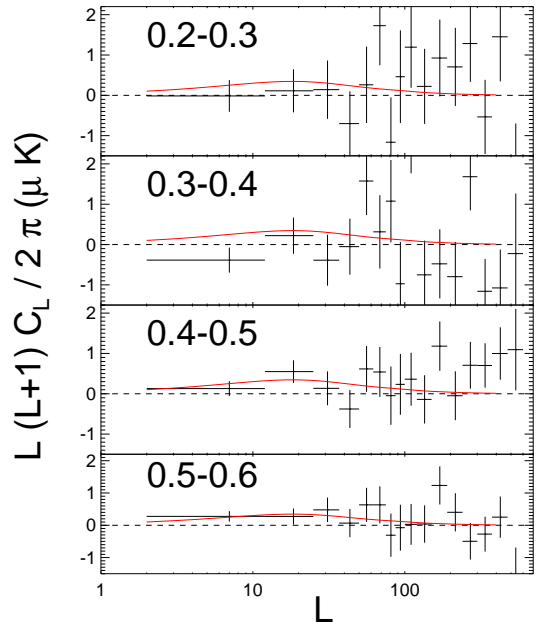}
\caption{\label{fig:isw_flat_zslice} The cross correlation between the 
\WMAP V band and the LRG density, measured in $\Delta z=0.1$ photometric redshift
slices. The bands are  exactly those used in Fig.~\ref{fig:isw_flatpower}.
Note that the power spectra are consistent within the estimated errors. Also plotted
is the ISW template scaled by the measured V band bias.}
\end{figure}

The measured cross correlation (Fig.~\ref{fig:isw_flatpower}) 
shows a marginally significant 
($\sim2\sigma$) excess at multipoles around $l \sim 50$, $l \sim 100$ and 
$l \sim 200$. This could either be a statistical fluctuation, or it
could represent foreground contamination. As discussed above, there
are a number of possible systematics -- we consider each of these in turn 
below.
\begin{itemize}
\item Galactic contamination : Given the size of this effect and the expected
contamination on these scales (Fig.~\ref{fig:isw_fgplot}), this 
would either require an order of magnitude mis-estimate of the 
known foregrounds, or a new component. Combined with the fact that this 
would have to be significant at high Galactic latitudes only (since 
the known templates agree well with the measured microwave emission in all
regions outside the Kp0 mask), 
Galactic contamination is unlikely to be the source of this power.
\item Extragalactic foregrounds : Extragalactic foregrounds broadly divide
into two classes -- SZ contamination and microwave point sources.
Known point sources have very strong frequency dependent spectra, not
observed in Fig.~\ref{fig:isw_flatpower}. The SZ effect is characterized by
a temperature decrement \cite{1999coph.book.....P}
at these frequencies, the opposite of what is observed. The kinetic SZ 
effect, although frequency independent, is significantly smaller than the
thermal SZ effect and should be negligible.
\item Underestimation of Errors : A third possibility is that the errors 
(derived from the Fisher matrix) are underestimated. 
However, comparing with simulations, we find that the Fisher errors agree to 
better than $\sim 10\%$. 
\end{itemize}

Since known systematics are unlikely to cause this excess power, we divide the 
LRG catalog into various subsamples to attempt to localize this effect. 
Fig.~\ref{fig:iswcaps} shows the cross-correlation with the two contiguous regions
of the SDSS (see Fig.~\ref{fig:lrgmask}), while Fig.~\ref{fig:isw_flat_zslice}
cross-correlates different photometric redshift slices. Within the errors, these
different subsamples are consistent with each other around $l \sim 50$; the 
redshift bin from $z=0.3$ to $0.4$  shows an excess of power at $l \sim 100$.
While this excess might be caused by a failure of the LRG selection algorithm
at the junction of Cut~I and Cut~II, a precise mechanism that 
would cause an excess in power when correlating with the CMB is not apparent.

At the current level of accuracy, it is impossible to determine
if this excess is simply statistical, or if it indicates systematic effects.
Greater sky coverage as the SDSS nears completion could help
answer this question. We note that this excess signal adds 
to the best fitted value of ISW amplitude in our analysis (compare
columns 2 and 3 of Table~\ref{tab:bias}), but at a small level since these
small scale modes are downweighted in the theoretical templates 
(Fig.~\ref{fig:isw_plottemplate}). This differs from other analyses
where this procedure was not adopted \cite{2003astro.ph..7335S}. 
In addition, most of previous work has been based on correlation function 
analyses, where correlations between the bins in the 
correlation function make a scale dependence of the signal 
difficult to identify.

\section{Cosmological constraints}
\label{sec:cosmology}

\begin{figure}
\includegraphics[width=3in]{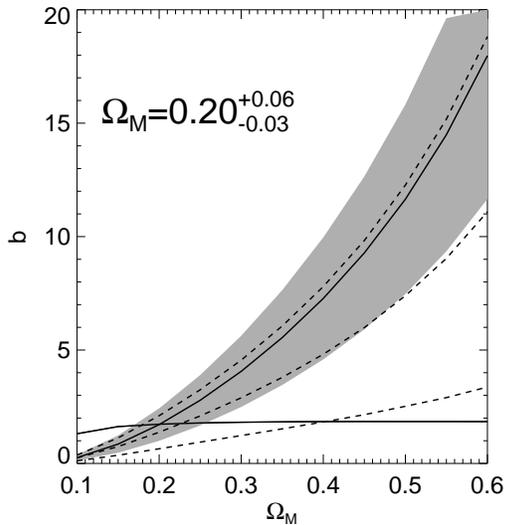}
\caption{\label{fig:isw_cosmo} The estimated galaxy bias from the
galaxy autocorrelation compared with the bias (and error) estimated from the ISW
signal (shaded region), as a function of $\Omega_{M}$ (assuming 
$\Omega_{M}+\Omega_{\Lambda}=1$). The errors on the autocorrelation bias are negligible
compared to the ISW errors. The best fit value of $\Omega_{M}$ is 
$\Omega_{M}=0.20^{+0.06}_{-0.03} (1\sigma) ^{+0.19}_{-0.05} (2\sigma)$. 
Note that the likelihood for $\Omega_{M}$ is extremely 
non-Gaussian (see Table~\ref{tab:cosmo}).}
\end{figure}

The fact that the ISW signal $\rightarrow \infty$ as $\Omega_{M} \rightarrow
0$ suggests a method to place a lower bound on the matter density within
the context of models with $\Omega_{M} + \Omega_{\Lambda} = 1$. If one computes
the bias both from the galaxy autocorrelation function ($b_{g}$) and 
the ISW effect ($b_{ISW}$), then one finds that $b_{ISW} \rightarrow 0$ as 
$\Omega_{M} \rightarrow 0$ faster than $b_{g}$, as does the error on $b_{ISW}$.
Comparing the two values of the bias allows one to constrain the 
value of $\Omega_{M}$. Note that this only puts a strong lower bound on 
$\Omega_{M}$. The upper bound is determined by how strongly one detects the ISW
effect; $\Omega_{M} = 1$ implies no ISW effect. Therefore, one would only expect
to be able to rule out $\Omega_{M}=1$ at $\sim 2.5 \sigma$, our quoted detection 
significance.

\begin{table*}
\begin{tabular}{cccccccccccc}
\hline\hline
$\Omega_{M}$ & $b_{g}$ &  
$b_{ISW}$ & $\sigma_{b,ISW}$ & $[\sigma]$ & $[\sigma] $ & $[\sigma]$ &
$b_{ISW}$ & $\sigma_{b,ISW}$ & $[\sigma]$ & $[\sigma] $ & $[\sigma]$ \\
& & $(l_{max}=50)$ & & $(\sigma_{8}=0.9)$ & $(\sigma_{8}=0.8)$ & $(\sigma_{8}=1.0)$ &
$(l_{max}=400)$ & & $(\sigma_{8}=0.9)$ & $(\sigma_{8}=0.8)$ & $(\sigma_{8}=1.0)$ \\
\hline
  0.10&  1.31&  0.24&  0.13&   8.3&   7.2&   9.5&  0.26&  0.13&   8.2&   7.1&   9.4\\
  0.15&  1.62&  0.74&  0.38&   2.3&   1.8&   2.8&  0.85&  0.38&   2.0&   1.6&   2.5\\
  0.20&  1.73&  1.38&  0.73&   0.5&   0.2&   0.8&  1.71&  0.71&   0.0&   0.2&   0.3\\
  0.25&  1.79&  2.09&  1.15&   0.3&   0.4&   0.1&  2.79&  1.11&   0.9&   1.1&   0.7\\
  0.30&  1.82&  2.89&  1.66&   0.6&   0.8&   0.5&  4.06&  1.57&   1.4&   1.6&   1.3\\
  0.35&  1.83&  3.79&  2.27&   0.9&   1.0&   0.8&  5.55&  2.09&   1.8&   1.9&   1.7\\
  0.40&  1.85&  4.81&  2.98&   1.0&   1.1&   0.9&  7.27&  2.69&   2.0&   2.1&   1.9\\
  0.45&  1.85&  5.99&  3.84&   1.1&   1.1&   1.0&  9.28&  3.37&   2.2&   2.3&   2.1\\
  0.50&  1.85&  7.39&  4.88&   1.1&   1.2&   1.1& 11.65&  4.18&   2.3&   2.4&   2.3\\
  0.55&  1.85&  9.04&  6.14&   1.2&   1.2&   1.1& 14.49&  5.14&   2.5&   2.5&   2.4\\
  0.60&  1.85& 11.09&  7.72&   1.2&   1.2&   1.2& 17.97&  6.32&   2.6&   2.6&   2.5\\

\hline
\end{tabular}
\caption{\label{tab:cosmo} The estimated bias from
the galaxy autocorrelation ($b_{g}$) and the ISW 
cross-correlation ($b_{ISW}$) as a function of $\Omega_{M}$, shown 
for both templates truncated at $l_{max}=50$ and $l_{max}=400$. 
The significance,$[\sigma]$, is defined as $|b_{g}-b_{ISW}|/\sigma_{b,ISW}$ where
$\sigma_{b,ISW}$ is the error on the bias inferred from the ISW effect, 
and $\sigma_{b,g} \ll 1$. This describes number of $\sigma$ a particular value of 
$\Omega_{M}$ is away from the
best fit value and can be used in likelihood analyses of 
cosmological parameters ($\chi^2=[\sigma]^2$). 
We also show the effect of changing
$\sigma_{8}$ on these results; increasing $\sigma_{8}$ increases the 
inferred value of $\Omega_{M}$. 
The ISW effect is seen to provide a strong lower bound on the value
of $\Omega_{M}$.}
\end{table*}

The results of performing this exercise are summarized in Fig.~\ref{fig:isw_cosmo}
and Table~\ref{tab:cosmo}. The galaxy and ISW bias are estimated
by using their predicted templates from Sec.~\ref{sec:theory}. 
We have limited ourselves to the 
\WMAP V band both for computational convenience and 
because the effects of galactic foregrounds are minimum there. We 
continue to use the fits to the measured power spectra used in 
previous sections to compute $C^{-1}$ as these correctly measure the cosmic
variance of these data. 

The best fit value for the matter density is 
$\Omega_{M}=0.20$ with $1\sigma$ limits of $\Omega_{M}=0.17$ and $\Omega_{M}=0.26$.
The strength of the lower bound is evident from Table~\ref{tab:cosmo}, where we
see that $\Omega_{M}=0.15$ is ruled out at $\sim 2 \sigma$, while $\Omega_{M}=0.10$ is
ruled out at $> 8\sigma$. Modest increases in the ISW detection
significance will translate into stronger constraints. For instance, increasing the survey
area by a factor of 2 (e.g., with the SDSS area on completion) could 
rule out $\Omega_{M}=0.15$ at $>3 \sigma$ while an all sky survey at moderate redshift
could rule it out at $>6 \sigma$.
While these bounds will not replace traditional parameter estimates 
(for eg.\cite{2004astro.ph..7372S}), 
they will provide much needed independent tests of dark energy. 

\begin{figure}
\includegraphics[width=3in]{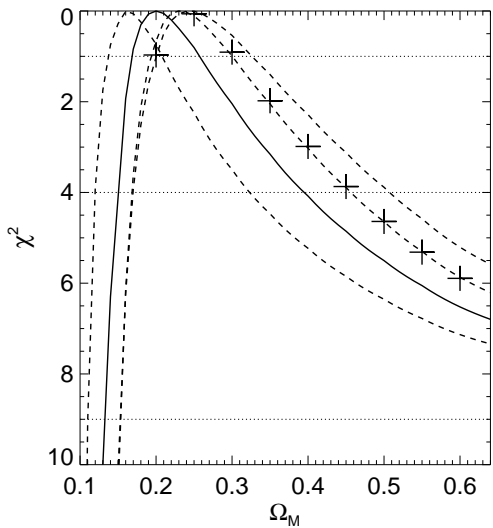}
\caption{\label{fig:isw_cosmoplot} The likelihood for $\Omega_{M}$ for different
values of the equation of state, $w$, computed using the prescription of 
Appendix~\ref{app:cosmo}. The solid line assumes $w=-1$, i.e. a cosmological
constant, while the dashed lines assume (from left to right) $w=-1.2$, $-0.8$, 
and $-0.5$ respectively. The crosses show the likelihood computed using the
full quadratic estimator discussed above, for $w=-0.8$, and test the approximations
made in the appendix. The dotted lines denote the 1-, 2-, and 3-$\sigma$ levels.}
\end{figure}

So far we have held all other cosmological parameters fixed in the 
above analysis. Marginalizing over these parameters, in general, will weaken
these constraints; these however will only weakly change the strength of the lower 
bound. For example, reducing $\sigma_{8}$ by a factor of $x$ increases the 
ISW bias by a factor of $x^{2}$, but the galaxy-galaxy bias only by a factor of 
$x$. However, decreasing $\sigma_{8}$ from $0.9$ to $0.8$ (a $10\%$
reduction) reduces the significance 
of $\Omega_{M}=0.15$ to $1.6\sigma$ from $2.0\sigma$, while 
making $\sigma_{8}=1$ increases it to $2.5 \sigma$. 

The above discussion assumes that the acceleration is due to a cosmological
constant, i.e. due to a component with equation of state, $P = -\rho$. An
important generalization are quintessence models with equation of state,
$P = w\rho$. For $w > -1$, the ISW signal is enhanced, increasing the best fit
value of $\Omega_{M}$ relative to the $w=-1$ case, while $w < -1$ reduces the 
signal, allowing for even lower values of $\Omega_{M}$; the likelihoods for 
$\Omega_{M}$ assuming different values of $w$ are shown in Fig.~\ref{fig:isw_cosmoplot}.
Unfortunately, the weak ISW detection does not, by itself,
allow us to place useful constraints on $\Omega_{M}$ and $w$ jointly. However, one 
can combine the ISW constraints with other data. This would 
constrain models with $w<-1$ that require a low $\Omega_{M}$, currently
disfavored by other data. We present an
simple algorithm for computing the likelihood 
for cosmological models parametrized by $(\Omega_{M}, \sigma_{8}, w)$
from the ISW signal in Appendix~\ref{app:cosmo}.

Finally, we make a parenthetical observation about the size of the errors 
for templates truncated at different $l_{max}$ values. We observe that for 
low $\Omega_{M}$ universes, truncating at different multipoles has little 
effect, while for higher $\Omega_{M}$, there is a significant difference. This is 
a direct result of the template weighting; at low $\Omega_{M}$, most of the 
weight is at low multipoles where the signal peaks and so, including 
higher multipoles has little effect. As $\Omega_{M}$ increases, the higher
multipoles gain importance, decreasing the error.

\section{Discussion}
\label{sec:discussion}

We have cross-correlated the microwave temperature maps observed by \WMAP with 
LRG overdensities obtained from the SDSS. 
Our goal is to extract the maximum amount of information on ISW 
available from the data. 
The major differences between 
our analysis and previous ISW analyses are:
\begin{itemize}
\item {\it Power Spectrum:} Most analyses (with the exception
of \cite{2004PhRvD..69h3524A}) have used the correlation function
instead of the power spectrum. While it is 
true that theoretically the power spectrum and correlation 
function are just Legendre
transforms of each other, this is no longer true in the presence of 
noise and sky cuts. 
In particular, the correlation function has extremely correlated
errors, while the power spectrum errors are almost uncorrelated on 
large scales. The presence of correlated errors makes an accurate
determination of the covariance matrix with either Monte Carlo 
or jack-knife techniques more difficult.
\item {\it Optimal $C^{-1}$ Weighting:} We weight our data with the
inverse covariance, resulting in  minimum-variance error bars
even in the presence of sky cuts. 
\item {\it Fisher Matrix errors:} We estimate our errors using the 
Fisher information matrix, instead of jack-knife error estimates favoured
by a number of other analyses. Jack-knife errors tend to underestimate
ISW errors because of the small number of uncorrelated jack-knife 
patches on the sky. These small numbers of samples also results in 
noisy covariance matrices, potentially biasing $\chi^{2}$ fitting.
We also note that the Fisher matrix errors are the smallest possible
errors, given the intrinsic sample variance. We verify that we saturate
this bound with simulations, indicating an optimal measurement.
  
\item {\it $\chi^{2}$ Fitting:} All analyses previous to this one
have estimated the significance of their detection by fitting
power spectrum or correlation function bins 
to ISW templates. We avoid this step by directly
estimating the amplitude of a fiducial template; the choice of this 
fiducial template affects our conclusions about the 
significance of the effect only marginally, as seen in Table~\ref{tab:cosmo}.
This has two advantages - (i) it optimally uses the information 
from all multipoles, and (ii) it avoids systematics introduced
by $\chi^{2}$ fitting with a noisy covariance matrix. The noise in the covariance
matrix is further exacerbated by strong correlations between bins; this 
results in badly conditioned covariance matrices where errors in the
most singular directions (the ones with the smallest eigenvalues) will dominate 
the fits. Note that these problems are made worse with the use of jack-knife 
errors, due to correlations in the different 
jack-knife samples on large scales and the small number of samples available.
\end{itemize}

The two analyses that use galaxies samples closest to ours are \cite{2003ApJ...597L..89F}
and \cite{2003astro.ph..7335S}; we note that both these analyses yield 
results similar to ours within the quoted errors. However, in addition to the
discussion above, our analyses differ in a number of areas. The most trivial is in the 
difference in the sky coverage, $\sim 4000~\rm{deg}^{2}$ compared to the $\sim 2000$ and
$\sim 3400~\rm{deg}^{2}$ respectively. Furthermore, we use accurately
calibrated photometric redshift distributions, including a deconvolution of the 
redshift errors \cite{2004astro.ph..7594P}. Most importantly, we use theoretical templates
to optimally weight the different scales, immunizing us to contamination
from statistical fluctuations on smaller scales.

Despite our attempt to be nearly optimal and our use of the latest and 
largest SDSS data sample we do not find a strong detection of ISW. 
We observe a correlation on 
large scales at 2.5 $\sigma$, consistent with the ISW effect. We fit to 
the amplitude of this effect, optimally combining the information from multipoles 
$< 400$ by using a predicted template, and, for $\Omega_M=0.3$ and 
$\sigma_8=0.9$, obtain a bias for the LRGs of 
$b_{g} = 4.05 \pm 1.56$ for the V band correlation, and similar
values for the other bands. Restricting to multipoles $<50$ yields
$b_{g} = 2.89 \pm 1.65$, consistent with the previous value although
with a somewhat larger error. These are
consistent with the value from the galaxy-galaxy 
auto-correlation, $b_{g} = 1.82 \pm 0.02$ and even better agreement 
is obtained for a smaller value of $\Omega_M$ $(\sim 0.2)$. 

We explore systematic
effects that could contaminate the signal. Known Galactic and
extragalactic foregrounds are subdominant to the measured
ISW signal. However, we do detect a marginally significant ($\sim 2\sigma$) 
correlation at scales of $l\sim 50$ and $l\sim100$. 
The frequency dependence and sign of the correlation
are inconsistent with extragalactic foregrounds, and Galactic foregrounds are 
approximately an order of magnitude too small to influence
our results. Subdividing the LRG sample
both on the sky, and in redshift slices, yields power spectra consistent with
each other, further ruling out systematic effects. We tentatively conclude
that this excess power is a statistical fluctuation, and wait for future data
to confirm or disprove this conclusion. This small scale excess power 
slightly increases the statistical significance of ISW in our analysis, 
but less so than in previous analyses where theoretical template 
weighting was not used. 

The ISW effect provides a useful and independent probe of the properties 
of the dark energy. Although the weakness of the current detections 
preclude detailed parameter estimations, we attempt to estimate the matter
density, by comparing the ISW bias to the value obtained from the 
auto-power spectrum analysis. The rapidly increasing amplitude of the 
ISW signal with decreasing matter density provides a strong lower bound
on the matter density, ruling out 
$\Omega_{M}=0.15$ at $2 \sigma$. These represent
fits to $\Omega_{M}$ only and assume $\sigma_8=0.9$ and
$w=-1$\cite{2004astro.ph..7372S}. 
The current ISW detection is not significant enough to jointly constrain
$\Omega_{M}$ and $w$. However, these constraints can be incorporated into 
parameter estimation efforts; we provide a simple likelihood prescription
to do this. As an example, assuming $w=-1.3$ reduces the predicted ISW signal 
and the matter density must be decreased to match observations: 
we find the best fit value for matter density is reduced to $\Omega_m=0.14$
with a $2\sigma$ upper limit of $0.29$. Such a low value is 
in conflict with other determinations of the matter density 
(see e.g. \cite{2004astro.ph..7377M}) and as a result 
$w<-1$ models are strongly constrained. 

In the present analysis we have ignored any evolution 
of the bias with redshift, since the redshift distribution is relatively 
narrow. Furthermore, estimates of the bias from the galaxy-galaxy 
power spectrum for different redshift slices yields a bias of between 
$1.7$ and $1.9$; this variation is too small to affect our analysis given 
the current errors.
However, this could  be a more significant issue for 
heterogeneous catalogs like the X-ray background
and the NVSS radio catalog.

ISW detection efforts are still in their infancy, and
future observations should lead to a stronger detection. This  
will hopefully realize the full potential of using
the ISW effect as a probe of dark energy.

\acknowledgments

We  acknowledge useful discussions with 
Niayesh Afshordi, Steve Boughn, Joseph Hennawi, Yeong-Shang Loh,  
Mike Nolta and Lyman Page.

CH is supported by the National Aeronautics and Space Administration
(NASA) Graduate Student Researchers Program (GSRP), grant no. NTGT5-50383.
US is supported by a fellowship from the
David and Lucile Packard Foundation,
NASA grants NAG5-1993, NASA NAG5-11489 and NSF grant CAREER-0132953.
We acknowledge use of the Princeton University Astrophysics Department's
Beowulf cluster, made possible through NSF grant no. AST-0216105.

Some of the results in this paper have been derived using the {\sc
healpix} \cite{1999elss.conf...37G} package. We acknowledge the use of the
Legacy Archive for Microwave Background Data Analysis (LAMBDA) 
\endnote{URL: \texttt{http://lambda.gsfc.nasa.gov}}. Support
for LAMBDA is provided by the NASA Office of Space Science.

Funding for the creation and distribution of the SDSS Archive has been
provided by the Alfred P. Sloan Foundation, the Participating
Institutions, the National Aeronautics and Space Administration, the
National Science Foundation, the U.S. Department of Energy, the Japanese
Monbukagakusho, and the Max Planck Society. The SDSS Web site is {\slshape
http://www.sdss.org/}.

The SDSS is managed by the Astrophysical Research Consortium (ARC) for the
Participating Institutions. The Participating Institutions are The
University of Chicago, Fermilab, the Institute for Advanced Study, the
Japan Participation Group, The Johns Hopkins University, the Korean
Scientist Group, Los Alamos
National Laboratory, the Max-Planck-Institute for Astronomy (MPIA), the
Max-Planck-Institute for Astrophysics (MPA), New Mexico State University,
University of Pittsburgh, Princeton University, the United States Naval
Observatory, and the University of Washington.

\appendix

\section{Cosmological Parameter Estimation}
\label{app:cosmo}

We discuss a prescription to extend the results of Table~\ref{tab:cosmo} 
to compute the 
likelihood of a cosmological model parametrized by $(\Omega_{M}, \sigma_{8},
w)$. The likelihood is computed with a $\chi^{2}$ function, comparing the 
galaxy bias obtained from the autocorrelation and ISW analyses,
\be
\chi^{2} = \left[\frac{b_{g}(\Omega_{M}, \sigma_{8}, w) - b_{ISW}
  (\Omega_{M}, \sigma_{8}, w)}{\sigma_{b,ISW}(\Omega_{M}, \sigma_{8}, w)}\right]^{2}
\,\, .
\label{eq:isw_chi2}
\ee
As in Sec.~\ref{sec:cosmology}, 
we ignore the error on $b_{g}$, since it is subdominant to the error
on the ISW measurement.

Table~\ref{tab:cosmo} presents measurements of $b_{g}$, $b_{ISW}$, and $\sigma_{b, ISW}$ as
a function of $\Omega_{M}$, for $\sigma_{8}=0.9$, and $w=-1$. Extending the
results for different $\sigma_{8}$ is straightforward, as discussed in 
Sec.~\ref{sec:cosmology},
\bea
b_{g}(\Omega_{M}, \sigma_{8}, -1) = b_{g}(\Omega_{M}) \left(\frac{0.9}{\sigma_{8}}
\right)\,\,, \\
b_{ISW}(\Omega_{M}, \sigma_{8}, -1) = b_{ISW}(\Omega_{M}) \left(\frac{0.9}{\sigma_{8}}
\right)^{2}\,\,, \\
\sigma_{b, ISW}(\Omega_{M}, \sigma_{8}, -1) = \sigma_{b,ISW}
(\Omega_{M}) \left(\frac{0.9}{\sigma_{8}} \right)^{2}\,\,,
\eea
where the right hand side asssumes $\sigma_{8}=0.9$, and $w=-1$, i.e. the values in
Table~\ref{tab:cosmo}.

\begin{table}
\begin{tabular}{cccccc}
\hline\hline
$w$ & & & $w_{bias}$& $(\Omega_{M})$ & \\
&0.1 & 0.2 & 0.3 & 0.4 & 0.5 \\
\hline
 -1.50 &  0.15 &  0.15 &  0.15 &  0.15 &  0.16 \\
 -1.40 &  0.28 &  0.27 &  0.27 &  0.27 &  0.28 \\
 -1.30 &  0.42 &  0.41 &  0.41 &  0.42 &  0.42 \\
 -1.20 &  0.59 &  0.58 &  0.58 &  0.59 &  0.59 \\
 -1.10 &  0.79 &  0.77 &  0.78 &  0.78 &  0.78 \\
 -1.00 &  1.00 &  1.00 &  1.00 &  1.00 &  1.00 \\
 -0.90 &  1.23 &  1.25 &  1.25 &  1.24 &  1.24 \\
 -0.80 &  1.46 &  1.51 &  1.51 &  1.50 &  1.49 \\
 -0.70 &  1.67 &  1.77 &  1.77 &  1.76 &  1.74 \\
 -0.60 &  1.83 &  1.99 &  2.00 &  1.99 &  1.97 \\
 -0.50 &  1.91 &  2.13 &  2.16 &  2.16 &  2.15 \\

\hline
\end{tabular}
\caption{\label{tab:wbias} Approximate multiplicative factors that scale the 
ISW power spectrum as a function of $w$ and $\Omega_{M}$. The bias scales as $b_{ISW}(w) = 
b_{ISW}(w=-1)/w_{bias}$. Note that the $\Omega_{M}$ dependence is weak, and can be 
ignored at the level of accuracy required.}
\end{table}

\begin{figure}
\includegraphics[width=3in]{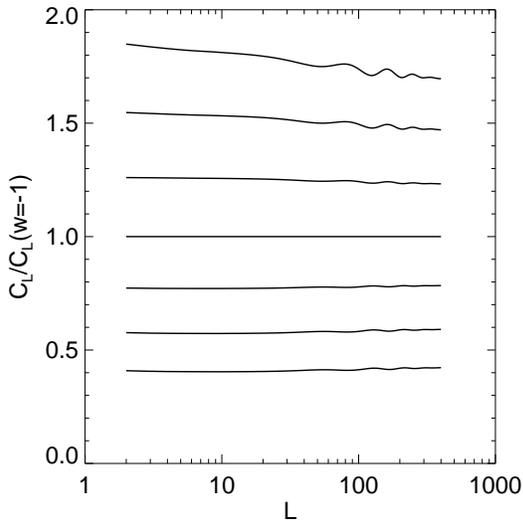}
\caption{\label{fig:isw_plot2_weq} The ISW angular power spectrum as a function 
of $w$, starting at the bottom with $w=-1.3$ and increasing in steps of $0.1$,
relative to $w=-1$. The cosmological parameters used are $\Omega_{M}=0.3$, 
$\Omega_{\Lambda}=0.7$, $\Omega_{b}=0.05$ and $h=0.7$. The power spectra can be
well approximated by a simple multiplicative bias.}
\end{figure}

The effect of changing $w$ on $b_{g}$ can be approximated by rescaling with the ratio of
the growth factor $D(z)$,
\be
b_{g}(\Omega_{M}, \sigma_{8}, w) = \frac{D(z_{mean}, w)}{D(z_{mean}, w=-1)} b_{g}(
\Omega_{M}, \sigma_{8}, -1) \,\,,
\ee
where $z_{mean} \sim 0.5$ is the median redshift of the LRG catalog. Computing the 
effect on $b_{ISW}$ would, in general, require a recomputation of the predicted ISW
signal; however, as Fig.~\ref{fig:isw_plot2_weq} indicates, 
we can approximate this (at the level of accuracy 
demanded by the strength of the ISW signal) by a simple multiplicative bias $w_{bias}$, 
tabulated in Table~\ref{tab:wbias}. The dependence of $w_{bias}$ is weak, and can be ignored. 
Therefore, we can complete our specification of $b_{ISW}$ and $\sigma_{b,ISW}$ with
\bea
b_{ISW}(\Omega_{M}, \sigma_{8}, w) = 
   \frac{b_{ISW}(\Omega_{M}, \sigma_{8}, -1)}{w_{bias}(w)} \,\,\\
\sigma_{b,ISW}(\Omega_{M},\sigma_{8}, w) = 
\frac{\sigma_{b,ISW}(\Omega_{M}, \sigma_{8}, -1)}{w_{bias}(w)}  
\,\,.
\eea

\bibliography{biblio,preprints}

\end{document}